\documentclass[prl,
				twocolumn,floatfix
				showpacs,
				superscriptaddress,
				amsfonts,amsmath,amssymb]{revtex4}

\usepackage{upgreek,xspace}
\usepackage[]{graphicx}

\newcommand{\mathcommand}[3][0]{\newcommand{#2}[#1]{\ensuremath{#3}}}
\mathcommand[1]{\ket}{\left| #1 \right\rangle}  
\mathcommand[1]{\bra}{\left\langle #1 \right|}  
\newcommand{\p}{$^{\text{31}}$P\xspace}
\newcommand{\D}{\ensuremath{\text{D}^\text{0}}\xspace}
\newcommand{\DX}{\ensuremath{\text{D}^\text{0}\text{X}}\xspace}
\newcommand{\mh}{\ensuremath{m_\text{h}}\xspace}
\newcommand{\sieight}{\ensuremath{{^{28}\text{Si}}}\xspace}

\newcommand{\sinine}{\ensuremath{{^{29}\text{Si}}}\xspace}

\newcommand{\downdown}{\ensuremath{\ket{\downarrow\Downarrow}}\xspace}
\newcommand{\updown}{\ensuremath{\ket{\uparrow\Downarrow}}\xspace}
\newcommand{\downup}{\ensuremath{\ket{\downarrow\Uparrow}}\xspace}
\newcommand{\upup}{\ensuremath{\ket{\uparrow\Uparrow}}\xspace}
\newcommand{\DtoDX}{\ensuremath{\D \rightarrow \DX}\xspace}

\begin{document}

\title{Simultaneous sub-second hyperpolarization of the
nuclear and electron spins of phosphorus in silicon by optical pumping of
exciton transitions}

\author{A.\ Yang}
\author{M.\ Steger}
\author{T.\ Sekiguchi}
\author{M.~L.~W.\ Thewalt}
\email{thewalt@sfu.ca}
\affiliation{Department of Physics, Simon Fraser University, Burnaby, British
Columbia, Canada V5A 1S6}

\author{T.~D.\ Ladd}
\affiliation{E.~L. Ginzton Laboratory, Stanford University, Stanford CA 94305, USA}

\author{K.~M.\ Itoh}
\affiliation{Keio University, Yokohama 223-8522, Japan}

\author{H. Riemann}
\author{N.~V. Abrosimov}
\affiliation{Institute for Crystal Growth (IKZ), 12489 Berlin, Germany}

\author{P. Becker}
\affiliation{PTB Braunschweig, 38116 Braunschweig, Germany}

\author{H.-J. Pohl}
\affiliation{VITCON Projectconsult GmbH, 07745 Jena, Germany}

\date{Accepted for publication in Physical Review Letters on June 03, 2009}

\begin{abstract}

We demonstrate a method which can hyperpolarize both the electron and nuclear
spins of \p donors in Si at low field, where both would be essentially
unpolarized in equilibrium. It is based on the selective ionization of donors in
a specific hyperfine state by optically pumping donor bound exciton hyperfine
transitions, which can be spectrally resolved in \sieight.  Electron and nuclear
polarizations of 90\% and 76\%, respectively, are obtained in less than a second,
providing an initialization mechanism for qubits based on these spins, and
enabling further ESR and NMR studies on dilute \p in \sieight.

\end{abstract}


\maketitle

Enriched \sieight is the material of choice for silicon-based quantum computing
schemes involving electron or nuclear
spins~\cite{Kane1998,Divincenzo2000,Vrijen2000,Morton2008,Ladd2002}, since the
removal of the \sinine nuclear spin results in very long coherence
times~\cite{Morton2008,Tyryshkin2003,Tyryshkin2006,Ladd2005}. Several methods for
achieving quantum logic with spin states of the shallow neutral donor (\D) \p in
\sieight have been proposed~\cite{Kane1998,Divincenzo2000,Vrijen2000} and the
manipulation of electron and nuclear spin coherences have been
demonstrated~\cite{Morton2008}, but unsolved challenges include the measurement
of single spins and the initialization, or polarization, of these spins. 
Fortuitously, the isotopic enrichment of \sieight has another dramatic effect:
the linewidths of many optical transitions are drastically
reduced~\cite{Karaiskaj2001,Karaiskaj2003,Thewalt2007,Yang2006,Thewalt2007a},
including those involving \p.  These narrow transitions have been proposed both for measurement
of single spins~\cite{Yang2006,Thewalt2007a,Fu2004} and for preferentially
populating specific spin states~\cite{Yang2006,Thewalt2007a}.

Electron and nuclear spin polarization in silicon has been studied for decades
\cite{Feher1956, Abragam1958,Feher1959a,Clark1963,Lampel1968,
Bagraev1982,Verhulst2005,Hayashi2006, Vlasenko2006,McCamey2009}, but the nuclear
polarization obtained to date has typically been less than a few percent, and
requires thousands of seconds to establish. Very recently, a \p nuclear
polarization of 68\% has been reported~\cite{McCamey2009} in a high magnetic
field, using a variation of a mechanism first proposed in 1959~\cite{Feher1959a},
and demonstrated in InSb in 1963~\cite{Clark1963}, but the time constant was
still a relatively long 150\,s. The method demonstrated here works at low
magnetic field, and can simultaneously hyperpolarize both the electron and
nuclear spins of \p in less than a second.
  
Preliminary attempts to demonstrate this mechanism yielded relatively small
electron and nuclear polarizations~\cite{Yang2009}, due to the fact that until
now all samples of \sieight with sufficiently high enrichment to resolve the
donor bound exciton (\DX) hyperfine transitions were $p$-type, with residual
boron acceptor concentrations typically ten times the \p concentration.  At low
temperature all donors were therefore ionized ($\text{D}^+$),  precluding the
observation of \DtoDX transitions, unless above-gap excitation provided
photoneutralization.  This need for above-gap excitation has a strong negative
effect on the achievable electron and nuclear polarizations, since it acts to
equalize the populations in the four \D hyperfine states.

The results presented here were made possible by a newly grown crystal of
\sieight purposely doped with \p.  This $n$-type sample allowed us to study the
\DtoDX transitions without requiring any above-gap excitation, resulting in
dramatically larger polarizations. An electron polarization of 90\% and nuclear
polarization of 76\% were obtained simultaneously in less than a second, in a
magnetic field and temperature regime where the equilibrium electron polarization
was only $\sim$2\%, and the nuclear polarization $\sim 3 \times 10^{-3}\%$. 
These large and rapidly produced hyperpolarizations should be sufficient for the
initialization of qubits, given the existence of algorithmic cooling techniques,
which expend modest qubit resources to quickly boost relatively high initial
polarizations above levels sufficient for fault-tolerant
operation~\cite{Schulman2005}.

The $n$-type \sieight sample central to obtaining these results was produced from
the same 99.991\% enriched \sieight as
previous~\cite{Yang2006,Thewalt2007a,Yang2009} $p$-type samples, except that
during the final floating zone growth of the dislocation-free single crystal, \p
was introduced using a dilute mixture of PH$_3$ in Ar carrier gas.  The resulting
crystal had a gradient in \p concentration, and the present results were obtained
from a slice containing $1 \times 10^{14}\,\text{cm}^{-3}$ boron and $7 \times
10^{14}\,\text{cm}^{-3}$ of \p. The sample was a disc with (001) faces, 1.5\,mm
thick and approximately 1\,cm in diameter, mounted in a completely strain free
manner in a reflecting cavity, and immersed in liquid He.  Other aspects of the
apparatus and methods have been described
previously~\cite{Yang2006,Thewalt2007a,Yang2009}.  The optical polarization of
the pump and probe lasers do not have a strong influence on the results.

It should be noted that these ultrahigh resolution \DX spectra cannot at present
be obtained via emission, or PL spectroscopy, the method typically used for
studies of \DX in semiconductors. The spectral resolution needed to resolve the
hyperfine splitting is beyond the capability of available spectrometers. Thanks
to tunable, single-frequency lasers, the needed high resolution can be achieved
in absorption mode, which we implement using photoluminescence excitation (PLE)
spectroscopy.

\begin{figure}[tb] \centering \includegraphics[width=1\columnwidth]{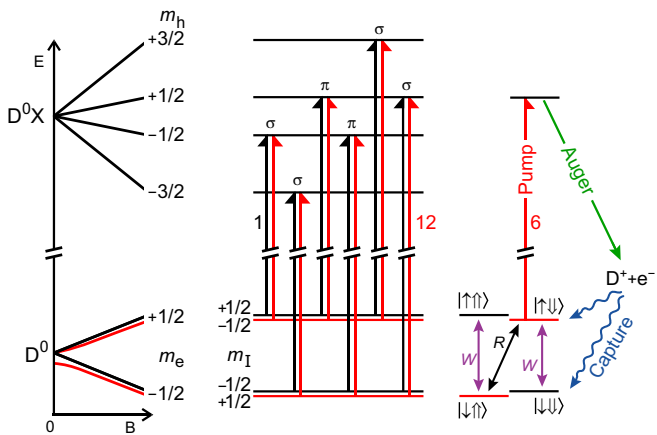}
\caption{\label{fig:leveldiagram}
(Color online) The neutral donor bound exciton transition and its low-field
Zeeman splittings are shown, along with a schematic of the selective optical polarization
mechanism. On the left are the splittings of the neutral donor ground state (\D)
and the bound exciton ground state (\DX). In the centre the twelve allowed
optical transitions between the four \D hyperfine states and the four \DX states
are indicated, numbered from 1 to 12 in order of increasing energy. On the right
is a simplified schematic of the optical polarization mechanism for 
the pump laser tuned to line 6.}
\end{figure}

The Zeeman spectroscopy of \p in \sieight in the low field regime is outlined in
FIG.~\ref{fig:leveldiagram}. On the left is a sketch of the splittings of the \D
and \DX ground states.  The Zeeman splitting of \DX is due primarily to the projection
of the hole angular momentum (\mh), since the two electrons form a spin singlet.
\D is split at zero magnetic field by the hyperfine interaction into an
electron-nuclear singlet and a triplet, separated by 117.53\,MHz
(486.1\,neV)~\cite{Feher1959}. Under a small magnetic field, these become two
hyperfine doublets, each separated by approximately one half of the zero-field
splitting.  The \D hyperfine states consist of two parallel spin states \upup
and \downdown which are not coupled by the hyperfine interaction, and two
anti-parallel states \updown and \downup which are somewhat mixed by the
hyperfine interaction.  The dipole allowed \DtoDX transitions are shown in the
centre of FIG.~\ref{fig:leveldiagram}, numbered from 1 to 12 in order of
increasing energy, and consist of six hyperfine-split doublets.  

\begin{figure}[tb] \centering \includegraphics[width=.8\columnwidth]{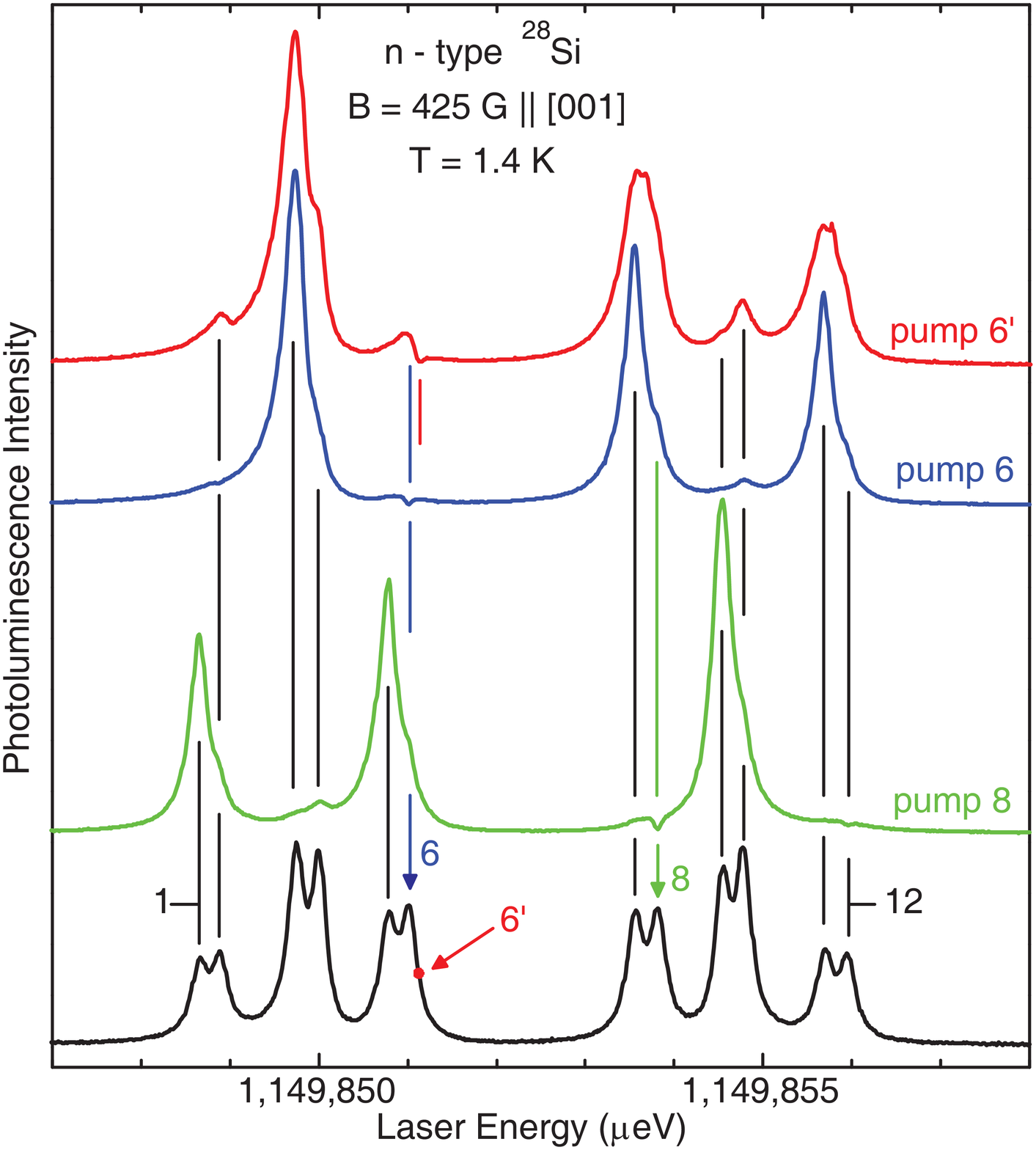}
\caption{\label{fig:optpumping}
(Color online) PLE spectra of the \p bound
exciton in an $n$-type \sieight sample, revealing electron and nuclear polarizations obtained by
selective optical pumping. At bottom is a spectrum without any resonant pumping,
showing essentially equal populations in the four \D states. The upper spectra
show the PLE signal when a strong pump field is tuned to either line 6 or line 8,
or the half-height point on the high energy side of line 6, which is indicated as
6$^\prime$.}
\end{figure}

The spectroscopic results are summarized in FIG.~\ref{fig:optpumping}, beginning
at the bottom with the unpumped spectrum of the $n$-type sample. This is very
similar to previous results~\cite{Yang2006,Thewalt2007a,Yang2009} for $p$-type
samples, except that the transitions are slightly less well resolved, with a full
width at half-maximum (FWHM) of  220\,neV (54\,MHz) as compared to 150\,neV
(37\,MHz).  Whether this small increase in linewidth is due to concentration
broadening from the \p, or to an unintentional reduction of the isotopic
enrichment, is not known at this time. The populations in the four \D hyperfine
states are seen to be essentially equal under these conditions, as expected.  It
is important to note that in order to obtain the spectrum at the bottom of
FIG.~\ref{fig:optpumping}, a small amount of above-gap excitation from a 1047\,nm
Nd-YLF laser was required in addition to the PLE probe laser.  This is not
because of a need for photoneutralization, as for $p$-type samples, but rather
because in the absence of any excitation other than the PLE probe laser, the PLE
probe laser itself strongly polarizes the \D hyperfine populations.  This pumping
effect of the probe laser results in a weak and distorted PLE spectrum, but these
saturation effects can be circumvented by repopulating the hyperfine states
equally with a small amount of above-gap excitation from the 1047\,nm laser.

For the other PLE spectra shown in FIG.~\ref{fig:optpumping}, a much stronger
pump laser ($\sim 2.8\,\text{Wcm}^{-2}$) is set to the desired energy while the
PLE probe laser ($\sim 6.5 \times 10^{-2}\,\text{Wcm}^{-2}$) is scanned across
the transitions. For the pumped spectra, no above-gap excitation is used, since
now the pump laser holds the populations of the hyperfine states essentially
fixed, and any above-gap excitation would merely reduce this desired
polarization. Unfortunately, without any above-gap excitation the PLE lines are
slightly broadened and develop a low energy tail.  This results from Stark
broadening due to the $\sim 1 \times 10^{14}\,\text{cm}^{-3}$ of ionized boron
and phosphorus which are present in the absence of above-gap excitation. This
broadening could be made negligible in a sample with a lower concentration of
boron.

The dramatic effects of selective pumping on the \D hyperfine populations are
immediately apparent in comparing the middle two spectra of
FIG.~\ref{fig:optpumping} to the unpumped spectrum. The results for pumping line
6 and 8 are shown since these produce the largest, and opposite, polarizations.
The resulting electronic polarization is extremely high, with transitions from \D
states having the same electron spin as the pumped state almost vanishing from
the spectra.  A large nuclear polarization in the hyperfine doublets from the
opposite electron spin states is also apparent. Identical results, except of
course for the energies of the transitions, were obtained when the magnetic field
was doubled and quadrupled. The top spectrum of FIG.~\ref{fig:optpumping} shows
the result when the pump laser was positioned at the half height point on the
high energy side of line 6, referred to as 6$^\prime$.  The results under the
conditions of FIG.~\ref{fig:optpumping} are summarized in
TABLE~\ref{tab:populations}, which gives the observed populations of the four \D
hyperfine states obtained by curve-fitting to the spectra, and the resulting net
electron and nuclear polarizations, when pumping at the peaks of lines 3 through
10 (pumping lines 1, 2, 11 and 12 give results very similar to lines 5, 6, 7 and
8, respectively).  Results essentially identical to these were obtained when the
pump and probe intensities were both reduced by a factor of one hundred,
indicating that high intensity is not needed to obtain high polarization.

The basic mechanism responsible for these polarizations is outlined on the right
hand side of FIG.~\ref{fig:leveldiagram} for the case of pumping line 6.  Only \D
in the \updown hyperfine state are converted into \DX, which decay with high
probability to ionized donors (D$^+$) plus free electrons (e$^-$), due to the 
dominance of Auger recombination~\cite{Schmid1977} for \DX in silicon. Subsequent
electron capture may then populate the opposite donor electron spin state.  The
pure nuclear relaxation rate is assumed to be negligible and therefore not shown,
and the pump rate is assumed to be much higher than either the optically enhanced
electron relaxation rate $W$, or the cross relaxation rate $R$.  Population is
therefore removed directly from \updown, and also from \upup and \downup,
since these are coupled to \updown via $R$ and $W$ relaxation, and builds up in
\downdown to the extent that the effective pump rate exceeds $W$.  Note that
this is very different from earlier mechanisms of nuclear polarization based on
either saturating ESR transitions~\cite{Feher1956} or on making
the $W$ and $R$ processes equilibrate at different
temperatures~\cite{Feher1959a,Clark1963,McCamey2009}. These earlier methods all
involve the bidirectional coupling of pairs of states, without or with a
Boltzmann factor, respectively, while selective optical pumping can move
population unidirectionally from the pumped state to the state having opposite
electron spin.

\begingroup
\begin{squeezetable}
\begin{table}[t]

\caption{\label{tab:populations}Observed \D populations and net electron and
nuclear polarizations derived from the spectra shown in
FIG.~\ref{fig:optpumping} and similar spectra for pumping lines 3, 4, 5, 7, 9
and~10.}
\begin{ruledtabular}
\begin{tabular}{r|rrrr|rr}
\multicolumn{1}{c|}{Pump}  & \multicolumn{4}{c|}{Populations (\%)} 
&\multicolumn{2}{c}{Polarization(\%)} \\ 

\multicolumn{1}{c|}{line} & \upup & \updown & \downdown & \downup & Elec. &
Nucl.\\ 
\hline 
3 \downdown & 44 & 38 & 4 & 14 & 64 & 16\\
4 \downup & 63 & 22 & 7 & 8 & 70 & 42\\
5 \upup & 1 & 8 & 75 & 16 & -82 & -66\\
6 \updown & 1 & 4 & 84 & 11 & -90 & -76\\
6$^\prime$ \updown & 2 & 11 & 64 & 23 & -74 & -50\\
7 \downdown & 64 & 26 & 1 & 9 & 80 & 46\\
8 \downup & 76 & 18 & 2 & 4 & 88 & 60\\
9 \upup & 3 & 15 & 43 & 39 & -64 & -16\\
10 \updown & 4 & 5 & 70 & 21 & -82 & -50\\

\end{tabular}
\end{ruledtabular}
\end{table}
\end{squeezetable}
\endgroup

The mechanism limiting the polarization is suggested by several observations.
When the pump and probe powers are reduced a factor of 100 the polarizations are
essentially unchanged. However when the pump laser is tuned off a peak the
polarizations are reduced substantially, as shown for 6$^\prime$ in
FIG.~\ref{fig:optpumping} and TABLE~\ref{tab:populations}. The achievable
polarization is therefore likely limited by nonselective photoionization of \D in
all four hyperfine states by the pump laser due to the tail of the \D
photoionization continuum.  Thus even larger hyperpolarizations could be achieved
if the substantial inhomogeneous broadening present in this sample could be
reduced.

A much more detailed model, incorporating the asymmetric inhomogeneous
broadening, the homogeneous linewidth, power broadening, and a dependence on the
hole state of the \DX has been constructed, which can reproduce the details of
the spectra, including the spectral hole burning evident at the pump laser energy
in the top three spectra in FIG.~\ref{fig:optpumping}.  A full account of the
model and the fits to various spectra is beyond the scope of this Letter, but one
important result is the fact that, independent of which component of a given
hyperfine doublet the pump laser is tuned to, the polarization results are
dominated by the coupling to the higher energy, or anti-parallel, component.

\begin{figure}[tb] \centering \includegraphics[width=.8\columnwidth]{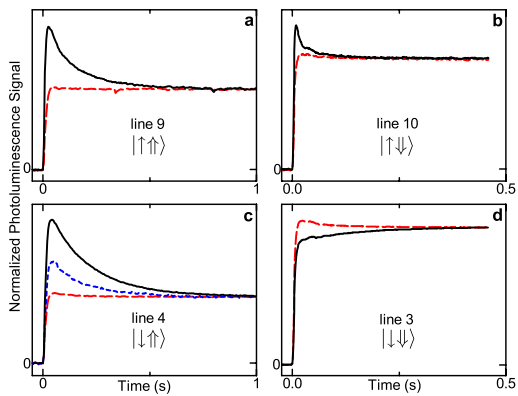}
\caption{\label{fig:transients}
(Color online) The transient behaviour of lines revealing the four \D
populations are shown, with the pump laser at line 6 in all cases.  Pump and probe are switched on
simultaneously at $t = 0$. The dashed curves show the transients with the
populations in the fully polarized state, while the solid curves show the
transients when the populations have been fully equilibrated. The middle curve in
\textbf{c} shows the transient after the fully polarized sample has recovered in
the dark for 20\,min.}
\end{figure}

To be useful for quantum computing and magnetic resonance, spin polarization must
occur on a reasonably fast timescale.  Due to the slow detector used in our PLE
apparatus, the system cannot capture the electron polarization dynamics, but is
still capable of following the time dependence of the nuclear polarization.
FIG.~\ref{fig:transients} shows the transient populations of the four \D
hyperfine states with the pump laser on line 6.  The transient when pump and
probe are turned on simultaneously is shown for both a fully polarized and a
fully depolarized initial state.  The transient from a polarized initial state is
a fast step function, but from the fully depolarized state
FIG.~\ref{fig:transients}\textbf{a}, \textbf{b}~and~\textbf{c} all show an
initial overshoot followed by a decay to the steady-state value, due to the
transfer of the hyperfine populations to the dominant \downdown state.
FIG.~\ref{fig:transients}\textbf{d} shows that the population in the \downdown
state, on the other hand, has a buildup following the initial fast transient,
with a $\sim$100\,ms time scale. FIG.~\ref{fig:transients}\textbf{c} also shows
the transient behaviour for line 4 when the fully saturated sample is allowed to
recover in the dark for 20\,min. The partial recovery of the unpolarized
population indicates that the nuclear spin relaxation time under these conditions
is 35\,min.

We have commented already on the fundamental difference between selective optical
pumping and previous methods used to obtain nuclear hyperpolarization.  The
hyperfine-resolved \DX transitions are also advantageous in being able to measure
directly the relative populations in all four hyperfine states, from which any of
the net polarizations or their correlations can be determined.  ESR measures
only the population difference between the \upup and \downup states, and between the
\updown and \downdown states, while NMR (which has not yet been feasible for
dilute \p in Si) measures the population differences between \upup and \updown,
and between \downup and \downdown states.  In analyzing their ESR data, \citeauthor{McCamey2009}~\cite{McCamey2009} have
assumed that the \upup vs. \downup and \updown vs. \downdown population
ratios are characterized by the same temperature, and that this temperature is the lattice
temperature.  The extent to which this assumption can be relied upon in the
presence of strong optical excitation is unclear.

The rapidly evolving nuclear polarization we observe is sufficiently high to be
useful for quantum computing. Our results also suggest that even higher
polarizations will be achieved as better samples of \sieight, with reduced
inhomogeneous broadening resulting from both higher isotopic enrichment and
reduced boron content, become available. Selective optical pumping may therefore
provide a viable initialization scheme for the \p nuclear spin as a qubit in
\sieight, and should work equally well for other shallow donors.  It should also
make possible NMR studies on \p in \sieight, both normal NMR using the
hyperpolarization, as well as optically detected NMR, revealing important new
information regarding this potential qubit. 

This work is supported by the Natural Sciences and Engineering Research Council
of Canada (NSERC).


\end{document}